# Large Current Modulation and Spin-Dependent Tunneling of Vertical Graphene/MoS$_2$ Heterostructures

*Nojoon Myoung[1], Kyungchul Seo[1,†], Seung Joo Lee[2,\*], and G Ihm[1,\*]*

[1]Department of Physics, Chungnam National University, Daejoen 305-764, Republic of Korea

[2]Quantum-functional Semiconductor Research Center, Dongguk University, Seoul 100-715, Republic of Korea

**ABSTRACT**: Vertical graphene heterostructures have been introduced as an alternative architecture for electronic devices by using quantum tunneling. Here, we present that the current on/off ratio of vertical graphene field-effect transistors is enhanced by using an armchair graphene nanoribbon as an electrode. Moreover, we report spin-dependent tunneling current of the graphene/MoS$_2$ heterostructures. When an atomically thin MoS$_2$ layer sandwiched between graphene electrodes becomes magnetic, Dirac fermions with different spins feel different height of the tunnel barrier, leading to spin-dependent tunneling. Our finding will develop the present graphene heterostructures for electronic devices by improving the device performance and by adding the possibility of spintronics based on graphene.

Graphene has been considered to be a promising material for future electronics due to its extraordinary properties such as high carrier mobility,[1,2] thermal conductivity,[3] and strong break strength.[4] Although the extremely high electrical conductivity makes graphene a potential candidate for replacing silicon-based electronics, Klein tunneling causes that electrical transport of Dirac fermions is insensitive to electrostatic potentials, resulting in a low current on/off ratio of graphene-based field-effect transistors.[5-7] In order to realize graphene electronics, it is important to manipulate its electronic properties without impairing the high mobility.

Recently, increasing interest has been focused on an alternative graphene device structure by using quantum tunneling. For a graphene/silicon heterojunction, large current on/off ratio was achieved by controlling the Schottky barrier formed at the interfaces.[8] In spite of the device performance, the carrier mobility of graphene deposited on silicon substrate is generally expected to decrease because of the inhomogeneity caused by the substrate.[9,10] Meanwhile, the possibility of a graphene field-effect transistor has been reported, based on vertical heterostructures with atomically thin insulating barriers such as hexagonal boron nitride (hBN) and molybdenum disulfide ($MoS_2$).[11-14] Layered materials such as hBN and $MoS_2$ have gained burgeoning interest as a material for use in graphene devices.[15] For example, the encapsulation of graphene by hBN maintains the high electronic quality of pristine graphene.[16-19] While the large band gap of hBN (~5.97 eV[20]) causes an insufficient current on/off ratio, the larger on/off ratio was observed for graphene/$MoS_2$ vertical field-effect transistor, owing to its smaller band gap as compared to hBN. Therefore, the graphene/$MoS_2$ heterostructure has been regarded as a significant building block of graphene-based electronics, and it is important to investigate the possible functional devices by utilizing its advantages for applications.

Herein, we present not only the improvement in the current on/off ratio of the existing graphene/$MoS_2$ vertical field-effect transistors[11] but also an application of the heterostructure in spintronics by producing spin-dependent tunneling. First, we show that there emerges the peak nature in tunneling current characteristics for a graphene/$MoS_2$/graphene nanoribbon (GNR) heterostructure. This finding has potential for the use of the current peaks, resulting in the improvement of the current on/off ratio. Second, the existence of magnetic properties in few-layer $MoS_2$[21-26] can lead to the spin-polarized current in the graphene heterostructures. We show that the graphene heterostruture can be a perfect spin-filter for holes with the electron-hole asymmetric spin splitting of $MoS_2$.[27] Such tunneling phenomena may lead to further potential applications in graphene-based electronics and spintronics.

Now, we consider a heterostructure, which consists of an atomically thin $MoS_2$ layer sandwiched between two graphene sheets as shown in Fig. 1a. It is well known that few-layer $MoS_2$ is an insulator with finite band gaps; ~1.9 eV direct band gap near K-valley and ~1.2 – 1.4 eV indirect band gap depending on the number of $MoS_2$ layers.[28-31] The $MoS_2$ layer of the heterostructure becomes a tunnel barrier for Dirac fermions, and both the graphene sheets play the role of high-quality source and drain electrodes. Dirac fermions experience the direct band gap near K-valley of $MoS_2$ rather than the smallest indirect band gap because of the momentum conservation, neglecting electron-phonon scattering processes.

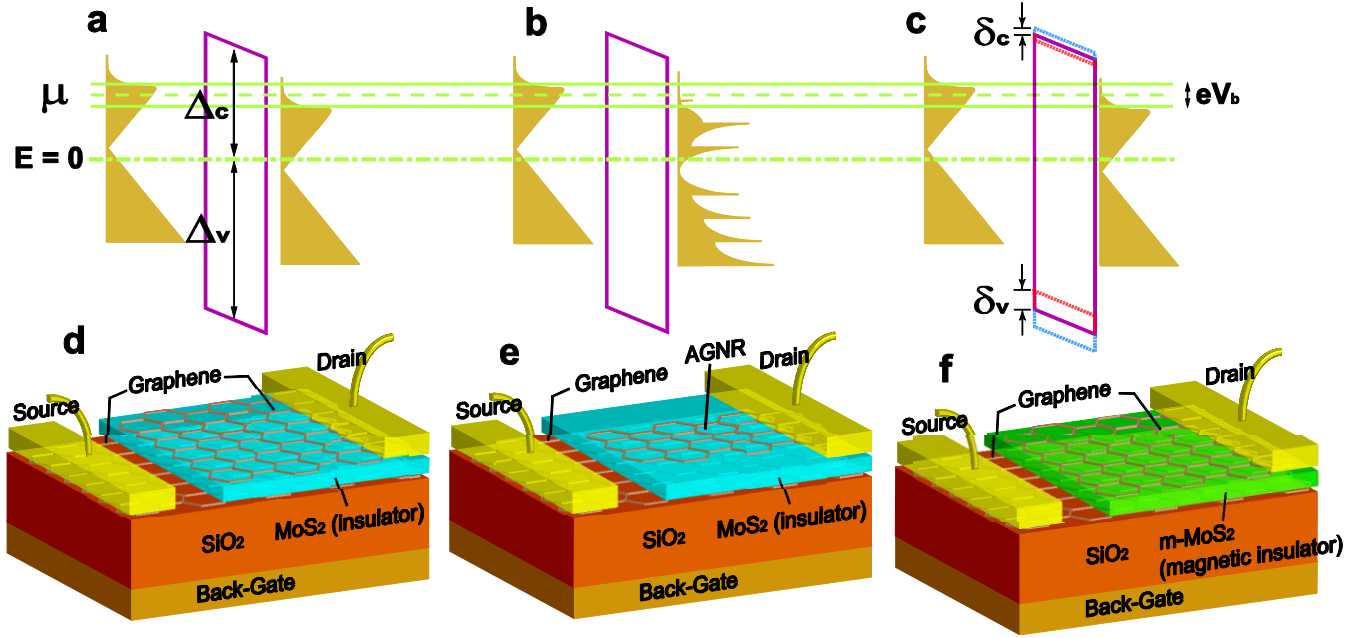

**Figure 1.** Schematics of various graphene/MoS$_2$ heterostructures considered in this study. (a), (b), and (c) Energetic diagrams for quantum tunneling through MoS$_2$ insulating barriers for various heterostructures; graphene/MoS$_2$/graphene, graphene/MoS$_2$/AGNR, and graphene/m-MoS$_2$/graphene, respectively. The chemical potential μ is formed by back gate voltage V$_g$, and the tunneling current is generated by bias voltage V$_g$ applied between source and drain graphene electrodes. Since the charge neutral point of graphene is asymmetrically laid, electrons and holes experience different tunnel barriers, Δ$_c$ and Δ$_v$. (b) If one graphene electrode is replaced by a narrow GNR, density of states is changed, reflecting the one-dimensional nature. (c) When the MoS$_2$ layer becomes magnetic, the barrier height is spin-dependent with different spin-splitting energies, δ$_c$ and δ$_v$. (d), (e), and (f) Schematic diagrams of various heterostructures corresponding to (a), (b), and (c), respectively.

### Results and Discussion

The tunneling current through the MoS$_2$ insulating barrier can be obtained as below

$$j(V_b, V_g) = j_0 \int_{-\infty}^{+\infty} D_s(E, V_b) D_d(E, V_b) T(E) [f_s(E, V_b, V_g) - f_d(E, V_b, V_g)] dE, \quad (1)$$

where $j_0 = (qv_F)/(2\pi L_0^2)$ is the unit of current density with electric charge of carriers q and the characteristic length of the system L$_0$. The transmission probability T(E) can be calculated quantum mechanically (see Supporting Information). Here, D$_i$ and f$_i$ are density of states of graphene and Fermi-Dirac distribution where i = s, d represent source and drain graphene electrodes on both sides of the MoS$_2$ layer, respectively.

By applying gate voltage $V_g$ *via* a back gate electrode, carriers are induced on top and bottom graphene layers. Simply, it can be assumed that the equal carrier concentration is induced on the both graphene layer; $n = \alpha V_g$ where $\alpha = 6.16 \times 10^{14}$ $C s^2 kg^{-1} m^{-4}$ in a case of 350 nm thick $SiO_2$ substrate. Under this assumption, chemical potential on both graphene layers are equally given as $\mu = \hbar v_F \sqrt{\pi |n|} = \hbar v_F \sqrt{\pi \alpha |V_g|}$ for the given $V_g$. In the absence of the bias voltage $V_b$ between top and bottom graphene layers, no net tunneling current produces. Applying $V_b$, one can measure non-zero tunneling current through the heterostructures. However, in fact, effects of the interlayer screening between top and bottom graphene layers must be taken into account in order to perform a much more detailed analysis of practical device performance. Since the interlayer screening length of graphene is short enough (~0.6 nm[32]), the bottom graphene layer, where carriers are induced by $V_g$, can affect the carrier concentration on the top graphene layer. For the given $V_g$, the carrier concentration induced on both graphene layers is calculated by solving Poisson's equation with inhomogeneous media (see Supporting Information). As a consequence of the screening by the bottom layer, always fewer carriers are induced in the top layer compared to the bottom layer. Due to this difference, Dirac cone of the top graphene layer shifts in order to bring the system in equilibrium.

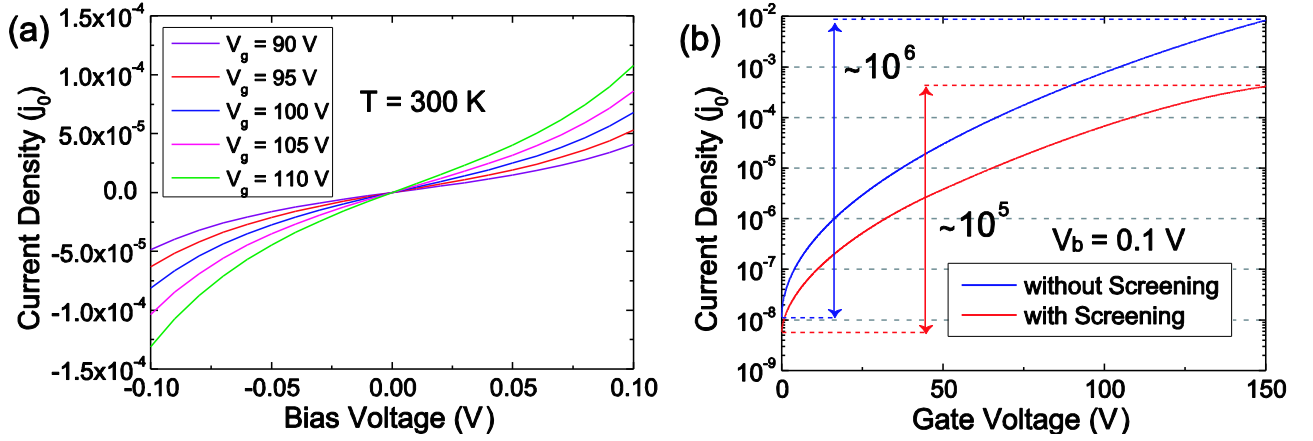

**Figure 2** Tunneling current characteristics of a graphene/MoS$_2$/graphene heterostructure. (a) Tunneling current density through graphene/MoS$_2$/graphene heterostructure as a function of bias voltage for different gate voltage at 300 K. (b) Tunneling current *versus* gate voltage for the given bias voltage with and without the interlayer screening between two graphene layers. The calculated tunneling current density is normalized by the current unit j$_0$ = (qv$_F$)/(2πL$_0^2$).

Figure 2 displays the characteristics of the graphene/MoS$_2$/graphene field-effect transistor. The calculated tunneling current as a function of V$_b$ for different V$_g$ are shown in Fig. 2(a). The tunneling current exhibits the increasing behavior with V$_b$ and becomes larger as V$_g$ increases. Here, note the fact that the tunneling current density is asymmetric with respect to bias voltage. This asymmetric feature of the tunneling current is a consequence of the interlayer screening. Even in equilibrium (V$_b$ = 0 V), there exists a finite electric field between the top and bottom graphene layers, which induces the shift of Dirac cone. When V$_b$ is applied between the two graphene layers, the number of carriers, which contribute to tunneling, is differently induced, depending upon the direction of V$_b$.

The tunneling current curve *versus* V$_g$ is plotted in Fig. 2(b). The ratio of the tunneling current density between an off-state (V$_g$ = 0 V) and an on-state (V$_g$ = 150 V) is found for the given V$_b$ at room temperature. While the current on/off ratio without the screening effect reaches up to 10$^6$, it goes down to 10$^5$ in the consideration of the screening effect. Despite of this decrease, the room-temperature current on/off ratio is still high as experimentally in Ref. 11.

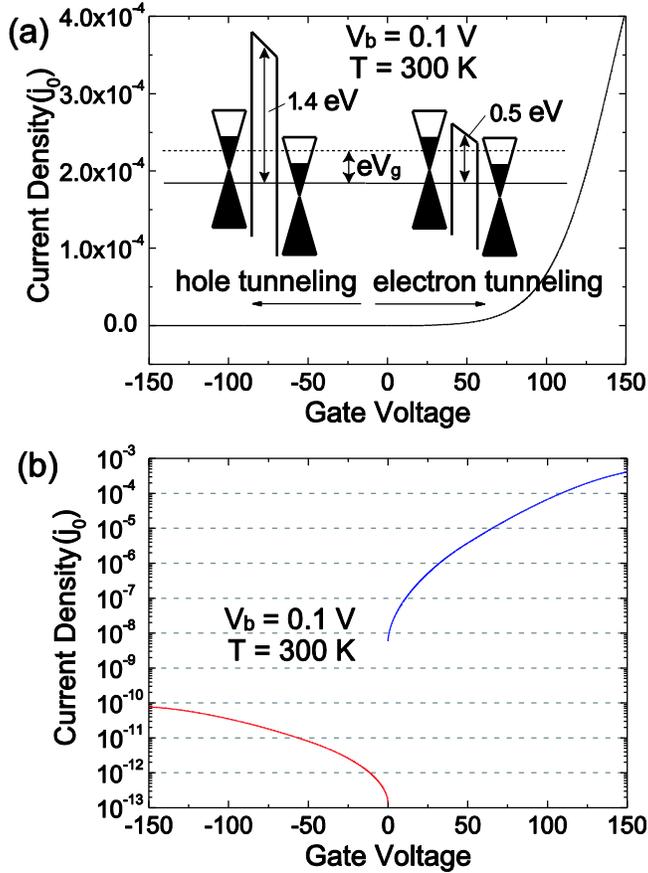

**Figure 3** Electron-hole asymmetry of tunneling current of a graphene/MoS$_2$/graphene heterostructure. (a) Asymmetric tunneling current density curve *versus* gate voltage at 300 K for V$_b$ = 0.1 V. Inset: Energetic diagrams for electron and hole tunneling. Electrons and holes experience different height of tunnel barriers because of the asymmetric arrangement of Dirac cone for graphene/MoS$_2$ hybrid system. (b) Linear-log plot of the electron and hole tunneling current density as a function of gate voltage. The current on/off ratio is also asymmetric for different kinds of carriers, electrons and holes.

For graphene/MoS$_2$ hybrid systems, the charge neutral point of Dirac cone is asymmetrically arranged between the conduction and valence bands near K-valley of MoS$_2$.[33] Thus, electrons experience a smaller tunnel barrier ($\Delta_c \approx 0.5$ eV) than holes ($\Delta_v \approx 1.4$ eV), where $\Delta_c$ and $\Delta_v$ are the barrier heights for electrons and holes, respectively. Therefore, if the tunneling electrons and holes are at the same chemical potential, the transmission probability through the MoS$_2$ insulating barrier for electrons is greater than for holes. Figure 3(a) exhibits the difference in tunneling current for electrons and holes as a function of V$_g$. We find that the tunneling current is indeed asymmetric with respect to the gate voltage polarity. The asymmetry also appears in the current on/off ratio as shown in Fig. 3(b). This result implies that it is advantageous to use electron tunneling for the vertical graphene/MoS$_2$ field-effect transistor. At this moment, let us note that the carrier-dependent tunneling current can become controllable by doping MoS$_2$ layers. (see Supporting Information)

Next, the current on/off ratio of the graphene/MoS$_2$/graphene field-effect transistor can be enhanced by considering finite size effects of a graphene electrode. A graphene electrode on one side of the sandwiched MoS$_2$ layer is replaced by a GNR as shown in Fig. 1(b). Since the tunneling current density depends on the product of density of states of source and drain electrodes (Eq. (1)), one can expect changes in the tunneling current characteristics. In this paper, armchair graphene nanoribbons (AGNRs) are considered, of which density of states exhibits the one-dimensional nature (see Supporting Information). Here, note that we can choose either armchair or zigzag graphene nanoribbons because the improvement in the current on/off ratio originates from the one-dimensional nature (van Hove singularity) of GNRs.

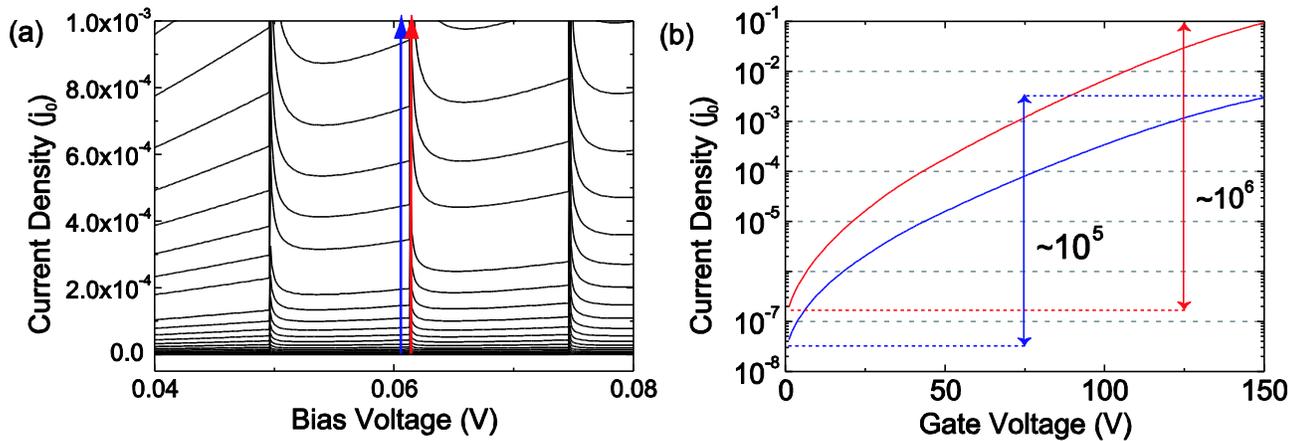

**Figure 4** Enhancement of current on/off ratio of a graphene/MoS$_2$/AGNR heterostructure. (a) Tunneling current density through graphene/MoS$_2$/AGNR heterostructures at 300 K as a function of bias voltage for different gate voltages. Peaks emerge in current density curves at specific $V_b$ due to the one-dimensional nature of AGNRs. (b) Comparison of the current on/off ratios in different cases; at peak (red lines) and near peak (blue lines), corresponding to Fig. 4(a)a. The calculated tunneling density is normalized by $j_0 = (qv_F)/(2\pi L_0^2)$.

The resulting tunneling current through a graphene/MoS2/AGNR heterostructure is plotted in Fig. 4(a) as a function of $V_b$ for different $V_g$, at 300 K. There emerge peaks in the tunneling current density curves at specific $V_b$ due to the existence of the van Hove singularities of the AGNR. This feature plays a crucial role in the enhancement of the current on/off ratio. Figure 4(b) shows linear-log plots of the tunneling current density as a function of $V_g$ for different $V_b$, at 300 K. The magnitude of the current density for $V_b = V_{peak}$ (at a current peak, $V_{peak} \approx 0.06135$ V, blue lines) is 10 times larger than the background values (for $V_b = 0.061$ V, red lines). Due to the existence of the current peak, there emerges an abrupt change in the tunneling current by one order of magnitude, and the resulting current on/off ratio is enhanced up to $10^6$ if one adjusts $V_b$ near $V_{peak}$. Therefore, the use of an AGNR instead of a two-dimensional graphene sheet as an electrode is attractive for applications in graphene-based electronics, improving the performance of the vertical graphene field-effect transistor.

In recent years, it has been found that the exhibited magnetic properties in an atomically thin MoS2 layer can be due to several causes; zigzag-terminated grain boundary[21-24] or sulfur-vacancy.[25] For example, the broken inversion symmetry due to the sulfur-vacancy leads to a splitting between different spin states for few-layer MoS2, whereas there is no spin-splitting for bulk MoS2.[27] In this paper, we consider that the MoS2 layer used in our heterostructure is thin enough to have a non-zero spin-splitting energy when inversion symmetry is broken. In this case, the thin MoS2 layer can be treated as a magnetic insulator with a spin-dependent barrier height, $U(z) = \Delta_{c,v} + \sigma \delta_{c,v} + qV_b z/d$ where $\sigma = \pm 1$ represents different spins and $\delta_{c,v}$ indicates spin-splitting energies in the conduction and valence bands of MoS2, respectively. Let us focus on the spin-splitting near K-valley of MoS2 because tunneling Dirac fermions experience the direct band gap near K-valley rather than the indirect gap as aforementioned.

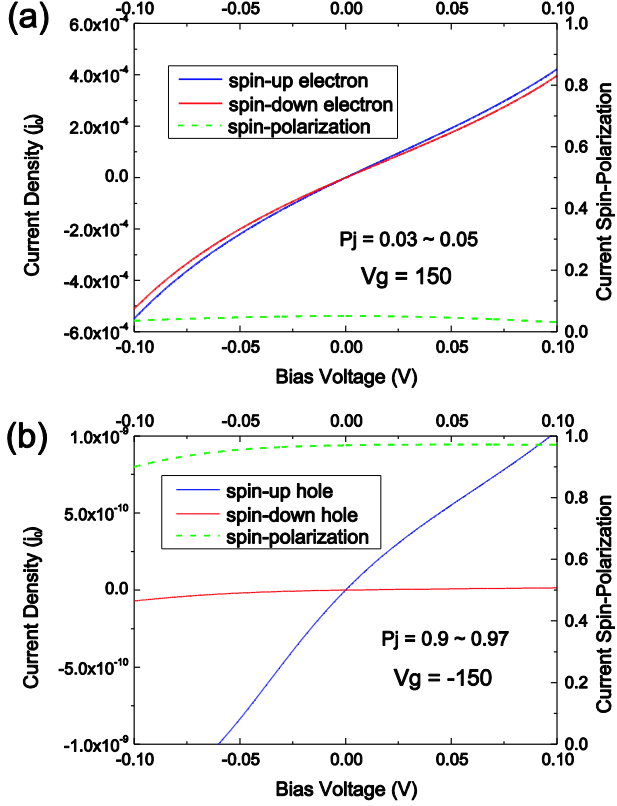

**Figure 5** Spin-polarization of tunneling current near K-valley. Spin-dependent current density through a graphene/m-MoS$_2$/graphene heterostructure for (a) electron and (b) hole tunneling. Blue and red solid lines indicate tunneling current densities for spin-up and down carriers, respectively. Green dashed lines represent the spin-polarization of the tunneling current density.

The calculation results of spin-dependent tunneling current through a magnetic MoS$_2$ (m-MoS$_2$) are shown in Fig. 5. Here, the spin-polarization of the tunneling current density is defined as $P_j = (j_{up} - j_{down})/(j_{up} + j_{down})$. In Fig. 5, one can see the differences in the spin-dependent feature for electron and hole tunneling currents. While the hole tunneling current is almost perfectly spin-polarized, the electron tunneling current is weakly spin-polarized. This is due to the fact that the spin-splitting near K-valley of m-MoS$_2$ is about 50 times larger in the valence band ($\delta_v \approx 145$ meV) than in the conduction band ($\delta_c \approx 3$ meV).[27] The small spin-splitting energy in the conduction band leads to the relatively weak spin-polarization of the electron tunneling current, $P_j \approx 0.03$-$0.05$. Meanwhile, due to the large spin-splitting in the valence band, the hole tunneling current is almost perfectly spin-polarized, $P_j \approx 0.9$-$0.97$. The spin-dependence of the tunneling current is also asymmetric with respect to the bias voltage polarity as consequence of the interlayer screening as aforementioned. Here, one may think that the large spin-polarization of the hole tunneling current seems to be unimportant because the hole tunneling is suppressed by the high tunnel barrier. This is resolved by using p-doped $MoS_2$ layers instead of intrinsic $MoS_2$. The hole tunneling current can be increased by p-doping of MoS2 layers as a consequence of reduction in the tunnel barrier height for holes. (see Supporting Information)

However, we need to consider the fact that there are two-equivalent valleys in $MoS_2$ (K and K') with the opposite spin-splitting energies. (These equivalent valleys in $MoS_2$ are significant because the opposite Berry curvatures at these valleys may affect on transport properties of $MoS_2$ such as the cancelation of in-plane current.[34]) In consequence, there should be the same amount of spin-up and spin-down Dirac fermions after tunneling through the spin-dependent tunnel barrier, and net current is non-spin-polarized in spite of the large spin-polarization near each valley. Here, let us introduce a strategy to achieve spin-polarized current in the graphene/m-$MoS_2$/graphene heterostructures; the valley-polarization in graphene electrode. Since most of spin-up(down) Dirac fermions in the drain graphene electrode is near K(K')-valley, we can achieve spin-polarized current if valley-polarization is generated by a valley-filter. It is well-known that the trigonal warping breaks the valley symmetry in graphene at several 100 meV, and it has been revealed that the valley polarization can be obtained through a p-n junction as a valley-filter.[35,36] (see Supporting Information) In our systems, chemical potential in the top graphene electrode is ~300 meV for $V_g$ = 100 V, which is valid for the trigonal warping. We, therefore, expect that the spin-polarized current can be achieved by adding a valley-filter to our heterostructure. In results, the graphene/m-$MoS_2$/graphene heterostructures provide a potential application in graphene-based spintronics as a good spin filter, compared with the existing spintronics technology.[37]

**Conclusions**

In summary, we have investigated the characteristics of the tunneling current through the graphene/$MoS_2$ heterostructures. We have done calculation for various heterostructures; graphene/$MoS_2$/graphene, graphene/$MoS_2$/AGNR, and graphene/m-$MoS_2$/graphene. It is shown that the current on/off ratio of the vertical graphene field-effect transistor based on graphene/$MoS_2$/graphene heterostructure is up to $10^5$ at room temperature. We have also found out that the vertical graphene field-effect transistor exhibits the carrier-dependent tunneling feature, which allows electrons to tunnel through the insulating barrier, more easily than holes. Furthermore, we have shown that the current on/off ratio of the vertical graphene field-effect transistor can be enhanced up to $10^6$ by replacing a two-dimensional graphene electrode with an AGNR electrode. Using a ZGNR electrode also leads to the enhancement of the current on/off ratio because the enhancement originates from van Hove singularity of GNRs. Finally, we propose a novel utility, the spin-polarized tunneling current, in the case that there exists a splitting between different spin states in the atomically thin $MoS_2$ layer. The different spin-splitting energy between the conduction and valence bands of $MoS_2$ makes the hole tunneling current almost perfectly spin-polarized whereas the electron tunneling current is partially spin-polarized. Although the hole tunneling is small due to the carrier-dependent tunneling, it may be resolved by further works on effects of p-doping of $MoS_2$ layers. The graphene/m-$MoS_2$/graphene heterostructure may acts as a spin-filter for holes, which can be a crucial building of future spintronic devices. Our findings not only offer an advance in research on the vertically stacked graphene heterostructures with thin insulating layers but also contribute to graphene-based electronic and spintronics.

**Method**

**Computational Details.** All tunneling current density data presented in this study were calculated numerically by using Fortran computer program (Compaq Visual Fortran, version 6.6, Compaq Computer Corp.). The Fortran 90 codes to perform the numerical calculations were developed by implementing own custom computational algorithm and subroutines from IMSL package. The calculation of the transmission probabilities through insulating barriers, which is necessary to obtain the tunneling current density, was done in direct tunneling regime. The curves of the non-linear carrier concentrations from the self-consistent Poisson's equation were performed in Mathematica (version 8.00, Wolfram Research, Inc.). The self-consistent problem was effectively solved by considering boundary conditions in static electrodynamics. For the graphene/$MoS_2$/AGNR heterostructures, density of states of AGNR is given as an analytic formula since AGNR exhibits exact solutions for its eigenmodes, but plots of the density of states shown in Supporting Information were obtained numerically by using Mathematica (version 8.00, Wolfram Research, Inc.).

*Conflict of Interest*: The authors declare no competing financial interest.

*Acknowledgment*: This work was supported by Basic Science Research Program through the NRF funded by the Ministry of Education, Science, and Technology (2012R1A1A4A01008299, NO2012R1A1A2005772).

*Supporting Information Available*: Detailed consideration of quantum tunneling through MoS$_2$ barriers, non-linearity of the carrier concentration with screening effect, current characteristics for doped MoS2, calculation of density of states of AGNRs, and consideration of trigonal warping in graphene. This material is available free of charge *via* the Internet at http://pubs.acs.org.

AUTHOR INFORMATION

**Corresponding Author**

*Email: (G.I.) ghihm@cnu.ac.kr; (S.J.L.) leesj@dongguk.edu

**Present Addresses**

† Daeduk Campus for BNT Convergence, Paichai University, Daejeon 305-509, Republic of Korea

**Author Contributions**

N.M. performed most of the calculations and mainly analyzed the results with in-depth discussions with K.S., S.J.L., and G.I. All the authors contributed to the writing of the manuscript.